# Bidirectional human-AI collaboration in brain tumour assessments improves both expert human and AI agent performance

James K. Ruffle[1,2,*], Samia Mohinta[1], Guilherme Pombo[3], Asthik Biswas[4,†], Alan Campbell[5,†], Indran Davagnanam[2,†], David Doig[2,†], Ahmed Hammam[2,†], Harpreet Hyare[1,2,6,†], Farrah Jabeen[2,7,†], Emma Lim[8,9,†], Dermot Mallon[2,†], Stephanie Owen[2,†], Sophie Wilkinson[6,†], Sebastian Brandner[1,10], Parashkev Nachev[1]

[1]Queen Square Institute of Neurology, University College London, London, UK
[2]Lysholm Department of Neuroradiology, National Hospital for Neurology and Neurosurgery, London, UK
[3]NVIDIA, UK.
[4]Great Ormond Street Hospital for Children, London, UK.
[5]Imaging Department, Royal National Orthopaedic Hospital, Stanmore, Middlesex, UK
[6]Department of Radiology, University College Hospitals NHS Foundation Trust, London, UK.
[7]Department of Radiology, Royal Free Hospital, London, UK.
[8]Department of Imaging, Imperial College Healthcare NHS Trust, London, UK
[9]Department of Brain Sciences, Imperial College London, London, UK
[10]Division of Neuropathology and Department of Neurodegenerative Disease, Queen Square Institute of Neurology, University College London, London, UK

*Correspondence to:
Dr James K Ruffle
Email: j.ruffle@ucl.ac.uk
Address: Institute of Neurology, UCL, London WC1N 3BG, UK

†These authors contributed equally.

## Funding
JKR is supported by the Medical Research Council (MR/X00046X/1 & UKRI1389), the British Society of Neuroradiology, and the European Society of Radiology (ESR) in collaboration with the European Institute for Biomedical Imaging Research (EIBIR). HH and JKR are supported by the National Brain Appeal. PN is supported by the Wellcome Trust (213038/Z/18/Z). JKR, PN and HH are supported by the UCLH NIHR Biomedical Research Centre.

## Conflict of interest
None to declare.

## Manuscript Type
Original article

## Authorship
JKR – led the study, conceptualisation, design, data acquisition, analysis, and manuscript drafting. SM, GP – assistance with data acquisition and analysis. AB, AC, ID, DD, AH, HH, FJ, EL, DM, SO, SW – radiology reviews, SB – data acquisition, PN – conceptualisation, infrastructural support, analysis, manuscript drafting.

All authors have contributed to the writing of the manuscript and have read and approved the final version.

## Abstract

The benefits of artificial intelligence (AI) human partnerships—evaluating how AI agents enhance expert human performance—are increasingly studied. Though rarely evaluated in healthcare, an inverse approach is possible: AI benefiting from the support of an expert human agent. Here, we investigate both human-AI clinical partnership paradigms in the magnetic resonance imaging-guided characterisation of patients with brain tumours. We reveal that human-AI partnerships improve accuracy and metacognitive ability not only for radiologists supported by AI, but also for AI agents supported by radiologists. Moreover, the greatest patient benefit was evident with an AI agent supported by a human one. Synergistic improvements in agent accuracy, metacognitive performance, and inter-rater agreement suggest that AI can create more capable, confident, and consistent clinical agents, whether human or model-based. Our work suggests that the maximal value of AI in healthcare could emerge not from replacing human intelligence, but from AI agents that routinely leverage and amplify it.

## Main

Integrating artificial intelligence (AI) systems into clinical practice represents one of the most promising technological advances for modern medicine[1-7]. But, whereas a great deal of initial research and development has focused on tools that might ultimately reduce, or even replace, clinician involvement in a patient's care[8-13], research demonstrating AI's capacity to enhance existing practice through AI-agent and human-agent partnerships is now actively growing[3,4,6,14-19]. Whilst many are illuminating in their field-advancing discoveries, it is notable that all of these studies adopt a unidirectional approach to human-AI partnership, namely, the human agent's performance gains in a clinical task when supported by an AI agent[3,4,6,14,17-19]. For example, Wu *et al.* recently conducted a randomised controlled trial comparing diagnostic accuracy among ophthalmologists with and without AI support, reporting superior accuracy when the model was used[17].

Although rarely evaluated in healthcare, an inverse paradigm is possible, where AI performance is guided by a human expert. (Extended Data Fig. 1). Similarities could be drawn from autopilot systems in both the aviation and car industries[20-22]. A vehicle may have driver-assistance features, such as lane-departure warnings, that activate when its human operator steers beyond a lane's boundaries. Alternatively, a vehicle may have autopilot systems that entirely control its path. Yet, it still contains a yoke/steering wheel, should human support of the pilot/driver be required (Extended Data Fig. 2). Providing the best possible care for the individual patient should always be our primary concern[23], but since this alternative human-AI partnership paradigm is understudied, we remain in the dark about whether the greatest patient benefits may come from clinical human agents supported by AI, or rather, from AI agents supported by human ones.

Our task is to evaluate these two alternative paradigms in a challenging healthcare setting. We examine the specialist evaluation of brain tumour imaging data, evaluating the benefits not only for the patient but also for the practitioner and healthcare provider. Brain tumours are heterogeneous in appearance, varying widely in size, shape, signal characteristics, and location throughout the brain, as assessed with magnetic resonance imaging (MRI)[24-28]. The MRI assessment will typically include (though it is highly centre-dependent) a variety of structural imaging sequences, including T1-weighted, T2-weighted, Fluid-Attenuated Inversion Recovery (FLAIR)-weighted, and post-contrast T1-weighted sequences[29] (Extended Data Fig. 3). Post-contrast imaging (mostly a post-contrast T1-weighted sequence) is acquired following the intravenous injection of gadolinium contrast medium. The post-contrast T1 often, though not always, yields particularly informative and clinically actionable data, such as for the neurosurgeon planning and undertaking resection[30,31] or the oncology team tracing the lesion for stereotactic radiotherapy[32]. Nonetheless, it is not always desirable (or feasible) to administer gadolinium. This may be due to contraindications such as allergy or renal impairment, or it may simply be less desirable, particularly in patients who undergo frequent follow-up imaging, including children[33-39].

Requiring a radiologist to decide whether to administer gadolinium and acquire post-contrast sequences is, however, a challenging task. Definitionally, it requires a best guess (or perhaps more appropriately labelled, a 'gamble') to be made by the clinician for whether, given the non-contrast imaging sequences, added clinical benefit will be gleaned with additional post-contrast data[40,41]. It should come as no surprise that guesswork and gambles should be avoided at all costs in healthcare[23], and especially in high-stakes settings such as neuro-oncology[25], where calculated data-driven decision-making ought always to be preferred. This naturally raises the possibility that AI-enabled systems might help here. Multiple groups have recently demonstrated that deep learning models can reasonably predict whether a patient's brain tumour scan will contain enhancing disease (typically seen only with post-contrast imaging) when reviewing the non-contrast sequences alone[24,42-47].

Here, we investigate the utility of such a tool in a multi-case, multi-reader, randomised crossover study of expert agent performance among expert radiologists who ordinarily provide frontline care across some of the UK's largest and most specialist hospitals. In the largest known study of its kind, spanning ten datasets, four countries, and five neuro-oncology disease categories, we evaluate both human agents', in the form of board-certified experienced radiologists, and an AI agent's ability to predict whether a patient's MRI will contain post-contrast-enhancing neuro-oncological disease when reviewing pre-contrast sequences alone, both in isolation and with the support of the other. We reveal that, in this difficult clinical task, both diagnostic fidelity and metacognitive performance—the relationship between reporting confidence and diagnostic accuracy—increase for both human and AI agents when supported by the other, with a close relationship to pre-existing human experience. Moreover, by illustrating that the greatest possible benefits to patients may be yielded not from the classical paradigm of a human agent supported by AI, but instead from an AI agent supported by humans, our findings advocate a regime change in how healthcare human-AI partnership systems might be best utilised.

# Results

## Neuro-oncology patient cohort

Drawing on an overall sample of 11089 patients with brain tumours, we studied a model's previously locked held-out test set of 1109 unique cases[42]. In these, age was available for 522 patients (47.1%), with a mean ± standard deviation of 55.2 ± 16.6 years. Sex was available for 573 patients (51.7%), 300 of whom were male, and 273 were female. This included 753 patients with presurgical glioma, 155 with postoperative glioma resection, 100 with meningioma, 70 with metastases, and 31 with paediatric gliomas. 397 cases were from the UK, 614 from the USA, 82 from the Netherlands, and 16 from Sub-Saharan Africa.

## Radiologist participants

Eleven consultant radiologists agreed to participate in the trial, whose routine day-to-day radiology services are provided across six large London-based hospitals: some provide highly specialist neuro-oncology services, while the remainder provide more general healthcare services. Cumulatively, these six sites care for more than 3.3 million unique patients each year, both within London and as both international and national referral hubs. We deliberately included staff from these major sites that provide frontline care to increase the robustness and generalizability of the findings. The extent of their clinical radiology experience ranged from 5 to 20 years, with a mean of 11.3 years (interquartile range [IQR] 7.5-14.5). From a case pool of 1109 unique patients, 564 were randomly selected for review by the 11 radiologists. With each radiologist reporting 100 cases, case overlap ranged from 2 to 18 cases for each radiologist-radiologist pair. The model also reviewed all cases.

## Patient-centric effects

Overall, the best agent-performance combinations (in descending order) were as follows: 1) model with radiologist support (balanced accuracy [BA] 0.841), 2) model alone (BA 0.824), 3) radiologist with model support (BA 0.743), and 4) radiologist alone (BA 0.698), revealing the trajectory that AI supported by human agent partnerships can achieve higher performances than current standard human with AI support paradigms (Fig. 1).

Radiologist accuracy, sensitivity, specificity, precision, and F1 score without model support was 0.698 (95% CI 0.658-0.739), 0.798 (95% CI 0.756-0.841), 0.515 (95% CI 0.429-0.600), 0.693 (95% CI 0.641-0.744), and 0.714 (95% CI 0.682-0.745), respectively. Radiologist accuracy, sensitivity, specificity, precision, and F1 score with model support was 0.743 (95% CI 0.711-0.775), 0.893 (95% CI 0.852-0.934), 0.567 (95% CI 0.520-0.615), 0.714 (95% CI 0.681-0.746), and 0.760 (95% CI 0.729-0.791), respectively. The mean percentage change in radiologist accuracy was 6.4%.

Model accuracy, sensitivity, specificity, precision, and F1 score without radiologist support was 0.824 (95% CI 0.774-0.874), 0.920 (95% CI 0.870-0.970), 0.727 (95% CI 0.677-0.777), 0.771 (95% CI 0.721-0.821), and 0.839 (95% CI 0.789-0.889), respectively. Model accuracy, sensitivity, specificity, precision, and F1 score with radiologist support was 0.841 (95% CI 0.791-0.891), 0.910 (95% CI 0.860-0.960), 0.779 (95% CI 0.729-0.829), 0.802 (95% CI 0.752-0.852), and 0.848 (95% CI 0.798-0.898), respectively. The mean percentage change in model accuracy was 2.1%. These performance increases were stable across both folds of the nested cross-validation, and with experimental repeats with different randomised seeds.

Of the 11 radiologists, 8 (73%) exhibited improved reporting accuracy with model assistance. In one, performance remained unchanged; in two, it declined slightly. Case examples where a radiologist correctly reviewed a case only with the model's support are provided in Fig. 2. Case examples where the model correctly reviewed a case only with a radiologist's support are provided in Fig. 3.

## Practitioner-centric effects

Radiologist and model inter-rater agreements across 2389 pairwise patient case comparisons (1289 radiologist-radiologist pairs, 1100 radiologist-model pairs) increased significantly from a Cohen's kappa of 0.338 without agent support to 0.484 with agent support ($p$<0.0001). 77% of agent pairwise inter-rater agreements were higher with support from the other, leaving 23% higher without it. Model-radiologist agreement without support of the other had a Cohen's kappa of 0.314, significantly rising to 0.482 with each other's support ($p$<0.001) (Extended Data Fig. 4). Radiologist-radiologist agreement without support of the model had a Cohen's kappa of 0.333, rising to 0.486 with model support ($p$<0.001).

Radiologist reporting confidence significantly increased from a mean ± SD of 6.88 ± 1.92 (Likert /10) without the model, to 7.23 ± 1.93 with model support (+0.35 confidence gain, $p$<0.0001). Model calibrated confidence also significantly improved from 8.20 ± 3.44 without radiologist support to 8.60 ± 2.43 with radiologist support (+0.40 confidence gain, $p$<0.0001).

Radiologist reporting time significantly decreased from a mean ± SD of 45.6s ± 41.4s required per case to 30.6s ± 30.2s required per case with model support ($p$<0.0001). Translated into radiological throughput, this equates to a change in radiologist hourly throughput from 79 ± 71.7 cases reviewed per hour without the model to 118 ± 116.4 cases reviewed per hour with it: a 49% increase in reporting efficiency. Model throughput was 878 cases per hour (4.10s per case), regardless of whether they were supported by radiologists or working in isolation. Extended Data Fig. 5 shows the breakdown in accuracy, confidence, and response time for each individual radiologist with and without model support.

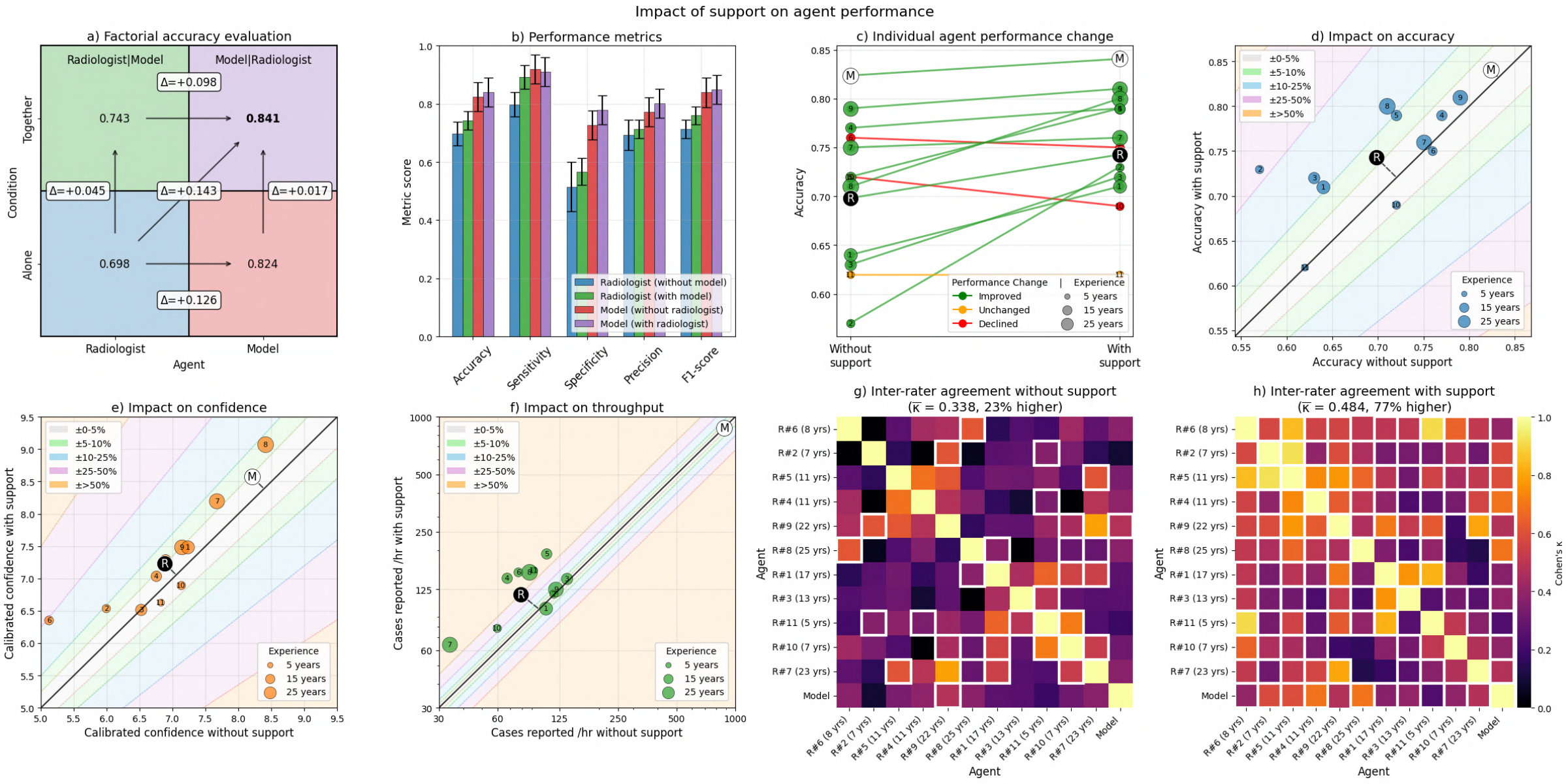

**Fig. 1. Impact of support on agent performance.** a) 2x2 factorial description of study paradigm, where human agents (trained radiologists) and an AI agent (trained model) were evaluated in their ability to detect post-contrast enhancing brain tumour using only non-contrast MRI, both alone and with the conditional support of the other. Balanced accuracy values are written within the quadrants, with delta differences related to the clinical status quo (radiologist alone, blue box). b) Metric comparisons for the radiologists alone (blue), radiologists with the support of the model (green), model alone (red), and model with radiologist support (purple). Error bars show bootstrapped 95% confidence intervals. c) Individual radiologist (as numbered within the scatter points) and AI agent performance with and without support of the other, colour-coded by performance change, sized by their years of experience, and the overall radiologist group 'R' effect, as well as the model 'M'. d) Impact of support on radiologist and model accuracy, e) confidence (Likert score /10 for radiologists, and calibrated confidence for the model), and f) throughput, showing each radiologist with points sized according to their experience, the radiologist group 'R' effect, and 'M' denoting the model. g-h) Agent inter-rater agreement with and without support of the other.

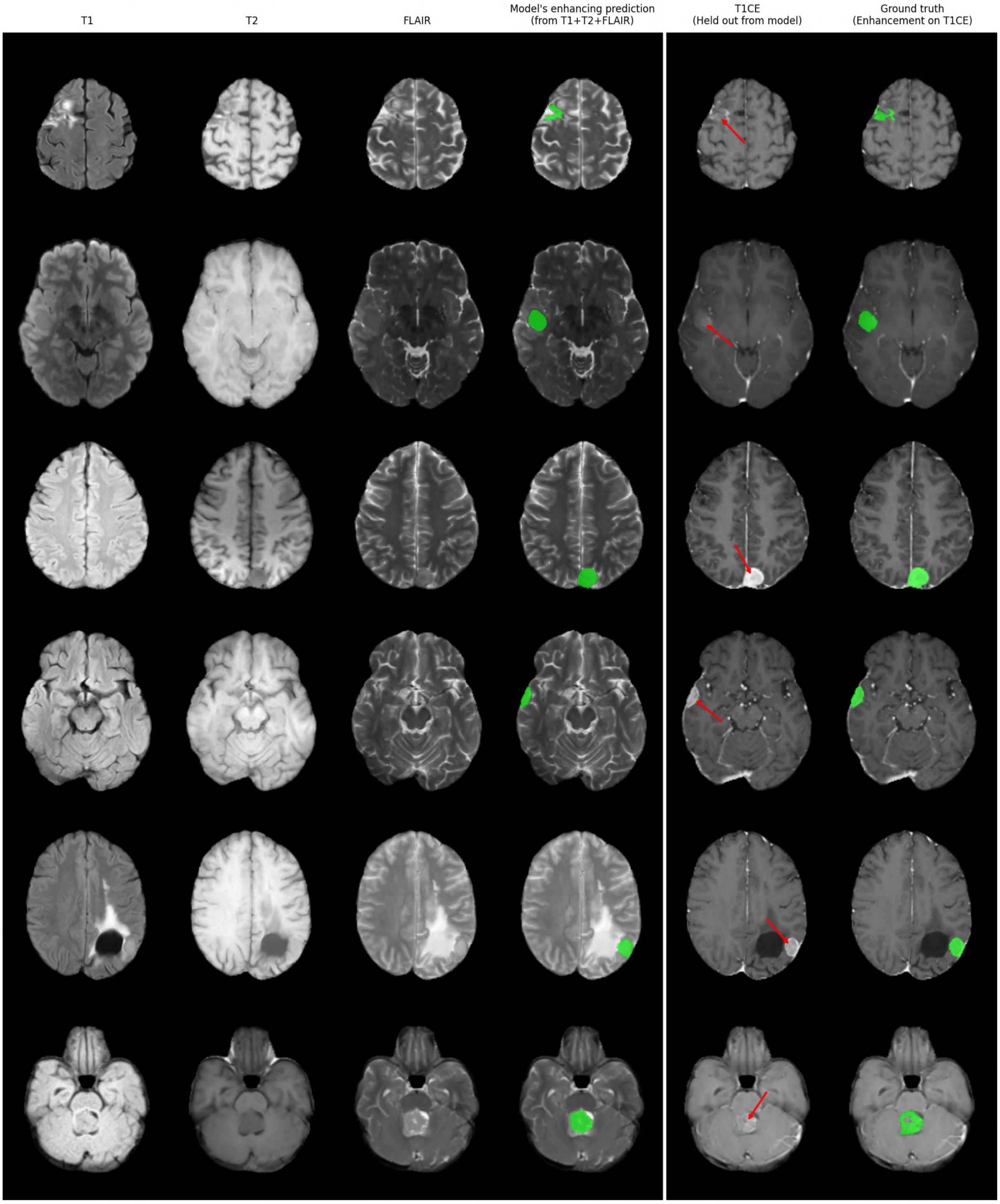

**Fig. 2. Enhancing patient cases where a radiologist's correct answer was only achieved with model support.** Rows show different patient cases, and columns 1-3 show the T1, T2, and FLAIR-weighted MRI sequences provided to both the radiologists and the model. The fourth column shows the model's prediction of post-contrast enhancement, based solely on those pre-contrast T1, T2, and FLAIR-weighted sequences. The fifth column shows the actual post-contrast T1 image (T1CE) with annotation of the contrast-enhancing disease (held out from both the model's and radiologist's review), and the sixth column shows the ground truth annotation (which was derived from the pre-contrast T1, T2, and FLAIR, and the post-contrast T1-weighted sequences).

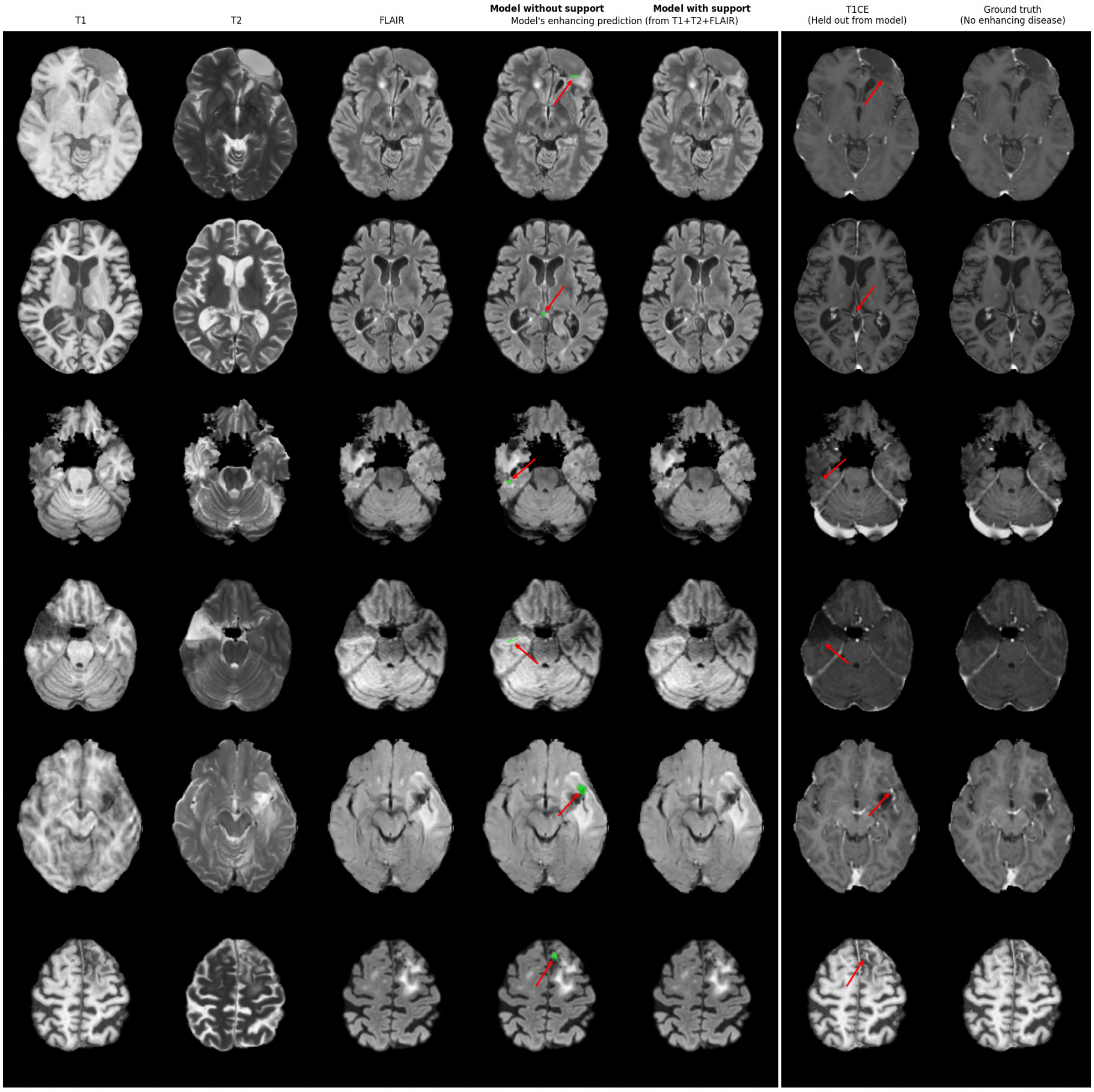

**Fig. 3. Nonenhancing patient cases where the model's correct answer was only achieved with radiologist support.** Rows show different patient cases, and columns 1-3 show the T1, T2, and FLAIR-weighted MRI sequences provided to both the radiologists and the model. The fourth column shows the model's initial prediction of lesional post-contrast enhancement based on the pre-contrast T1, T2 and FLAIR sequences. The fifth column displays the model's amended prediction after incorporating the radiologist's prediction and confidence. The sixth column shows the actual post-contrast T1 image (T1CE) (held out from both the model's and radiologist's review). The seventh column shows the ground truth annotation (derived from the pre-contrast T1, T2, and FLAIR, and the post-contrast T1-weighted sequences); empty here, as all cases shown did not contain post-contrast enhancing tumour. Arrows of the model prediction and ground truth images illustrate foci erroneously predicted to enhance by the model alone that were corrected following integration of human agent support.

The mean radiologist area under the receiver operator characteristic curve (AUROC) without model support was 0.729 (range 0.610-0.873) with an area under the precision recall curve (AUPRC) 0.724 (0.566-0.844), rising to an AUROC of 0.837 (range 0.759-0.909) and AUPRC of 0.832 (0.728-0.906) with the support of the model (Fig. 4). The model alone achieved an AUROC of 0.908 and AUPRC of 0.961, further rising with radiologist support to an AUROC of 0.914 and AUPRC of 0.976.

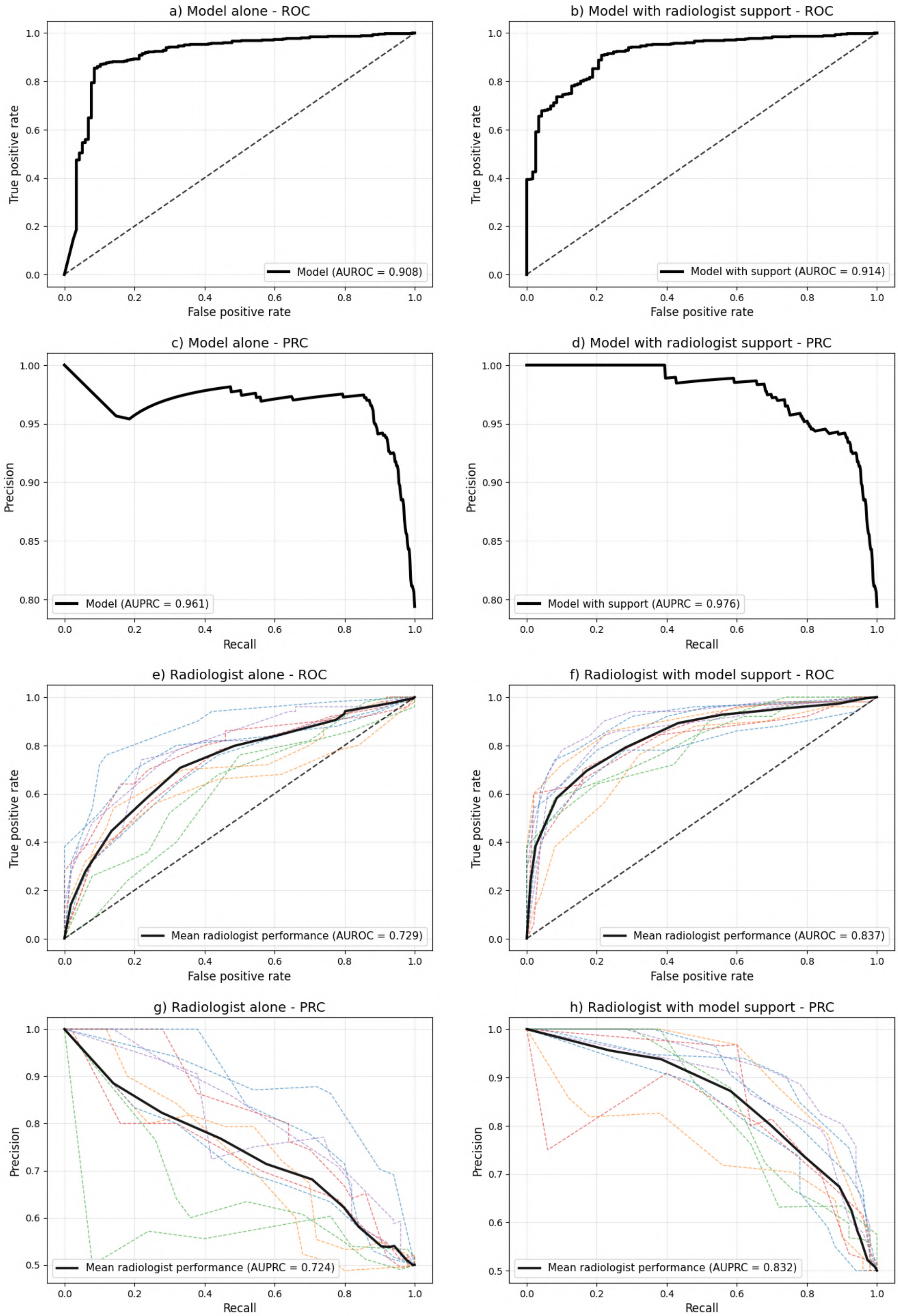

**Fig. 4. Model and radiologist performance comparison.** a-b) Model receiver operator characteristic (ROC) with and without radiologist support. c-d) Model precision-recall curve (PRC) with and without radiologist support. e-f) Radiologist ROC curve with and without model support. g-h) Radiologist PRC with and without model support. In panels e-h), coloured dashed lines show individual radiologists, with solid black lines showing the group mean. Abbreviations: AUROC, area under the receiver operator characteristic; AUPRC, area under the precision-recall curve.

Without model support, there was no statistically significant relationship between radiologists' years of experience and their accuracy ($R^2$ 0.156, $p$=0.230, Akaike Information Criterion [AIC] = -23.597) (Fig. 4). With model support, there was, however, a significant relationship between a radiologist's years of experience and their accuracy ($R^2$ 0.378, $p$<0.05), with an improvement in model fit criterion (AIC -32.174). A radiologist's years of experience were significantly related to their mean confidence, both without the model ($R^2$ 0.511, $p$<0.01) and were further strengthened with model support ($R^2$ 0.756, $p$<0.001, AIC without model 24.769, AIC with model 16.259). We fitted a function of $Pooled\ accuracy \sim Pooled\ confidence$, where a perfectly calibrated result would indicate a 0.100-unit increase in accuracy for every 1-unit increase in confidence. For radiologists without model support, a 1-unit increase in confidence score was associated with a 0.051 increase in accuracy (mean accuracy deviation of 0.170). In contrast, with model support, this improved to a mean unit increase of 0.105 (mean accuracy deviation of 0.089).

Without model support, there was a strong positive correlation between a radiologist's confidence (on a case-by-case basis) and their reporting speed (the number of cases reported per hour) (r 0.935, $p$<0.0001). With model support, the correlation between radiologist-reported case confidence and reporting speed was weaker (r 0.703, $p < 0.05$). Without model support, the radiologists' accuracy significantly correlated with their reporting speed (r 0.865, $p$<0.001). This relationship became non-significant with model support (r 0.563, $p$=0.09). Levene's test of variation demonstrated that reporting speed was significantly more variable across all cases, diseases, cohorts, datasets, and radiologists without model support, but it became significantly less variable with model support (coefficient of variation without model support: 64.1%, with model support: 50.4%) ($p$<0.01). Framed differently, the accuracy, confidence, and speed of radiologists working without model support were intricately related. However, with model support, this trilinear relationship was reduced, rendering reporting speeds similar regardless of case difficulty.

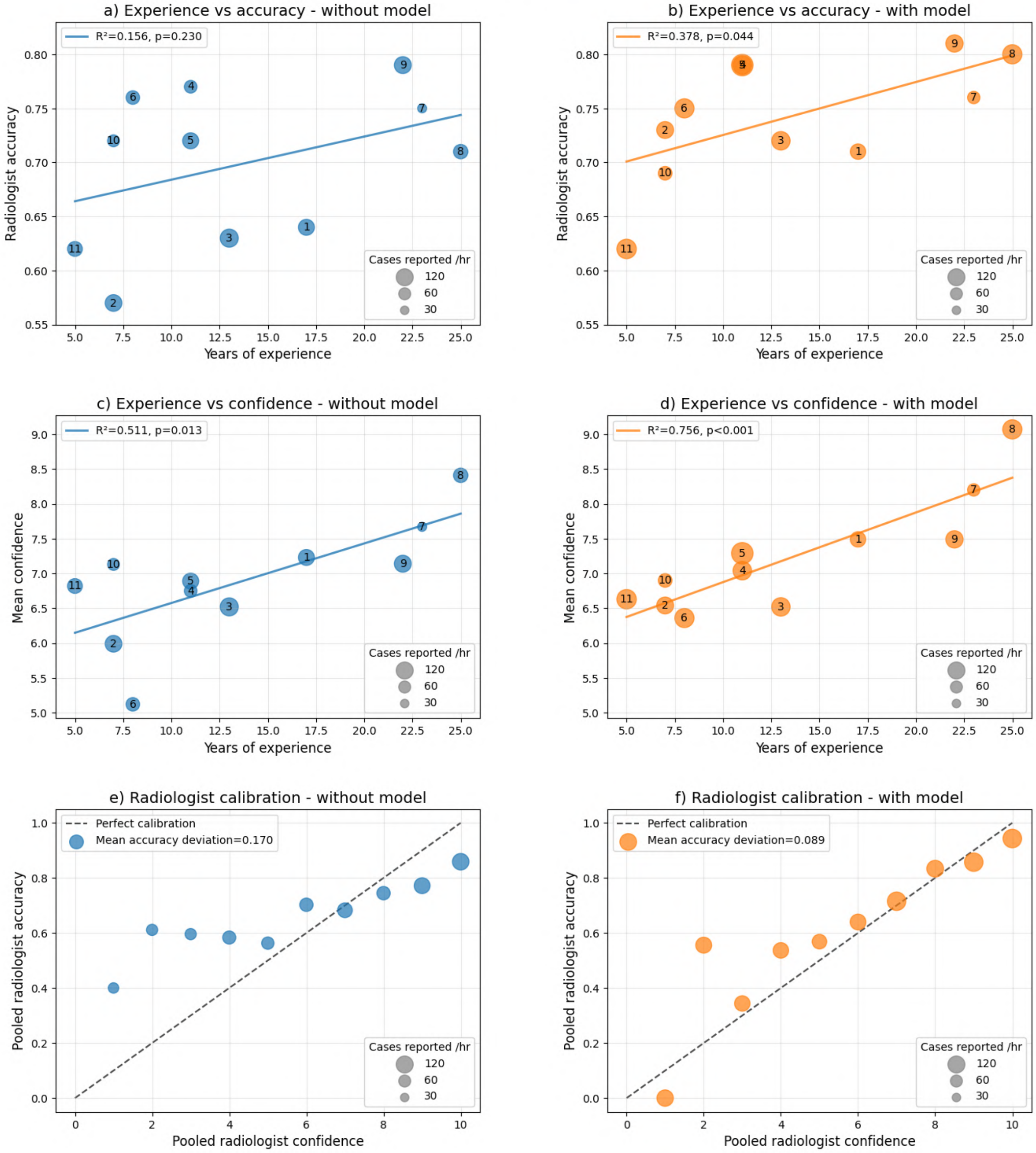

**Fig. 5. Strengthening the relationship between accuracy, experience, and confidence in radiologists.** a-b) Linear regression showing fit quality between a radiologist's years of experience and accuracy without (a) and with (b) model support. c-d) Linear regression showing fit quality between a radiologist's years of experience and reporting confidence without (c) and with (d) model support. e-f) $Pooled\ accuracy \sim Pooled\ confidence$ calibration function fit quality between radiologist confidence and binned accuracy without (e) and with (f) model support. Scatter points in panels a-d show individual radiologists (annotated with their identifier), and panels e-f show binned performances for each confidence value. Points are sized by the number of cases reported per hour.

Representation of ground truth masks into a compressed, two-dimensional UMAP-embedded space showed no clear relation between radiologist performance improvements across the high-dimensional representation of lesion morphology, pathology, size, and distribution (Extended Data Fig. 6). Radiologist accuracy and confidence increases were most notable for meningioma, metastases and presurgical glioma cases (Extended Data Fig. 7). Lesions between 0.5 cm³ - 10 cm³ were more accurately predicted in radiologists with model support, with more marginal performance changes for very large (>10 cm³) or tiny (<0.5 cm³) lesions. However, the largest increases in radiologist confidence were observed in large lesions (>5 cm³). Accuracy and confidence in radiologists with model support increased more notably in irregular, well-circumscribed, or infiltrative single lesions, while patient cases with multiple lesions were less affected. Reporting throughput increased among radiologists with model support, irrespective of pathology type, lesion size, or radiomic category.

## Healthcare system-centric effects

Total model training [experience] time was 2001.6 hours (83.4 days), 400.3 hours per fold (16.7 days), with an estimated training cost of £4804 under contemporary GPU hire charges (Fig. 6). In comparison, median radiologist experience time was 11 years (interquartile range [IQR] 7.5-19.5), or ~22,000 hours (IQR 15000-39000), assuming standard UK clinical job contracts. The median cumulative radiologist training/salary cost was £943,307 (IQR £549,053 - £2,013,008). With the addition of model support, the median additional radiologist experience leveraged by the model was 6.0 years (IQR 4.3 - 11.2), and the median staff financial value leveraged was £696,176 (IQR £557,885 - £1,458,254). This significantly increased median staff financial value to £2,668,256 (IQR £1,374,268 - £3,093,391), and median staff inferred experience to 24.3 years (IQR 14.6 - 27.3 years) (both $p<0.01$).

In further exploratory analysis, fitting the AI model to the radiologist experience-accuracy function inferred that the base model has an experience level similar to a radiologist with 45 years of clinical experience, with an anticipated experiential value of £5,566,135. With radiologist support, the model's inferred experience increased by 4 years to 49 years, resulting in a human agent-leveraged increase in model value of £559,528. This yielded a total accuracy-experience model value of £6,125,663. A breakdown of all radiologist and model performance experience, value confidence, and self-awareness changes following support from the other is provided in Fig. 6.

### Effect on practitioner calibration

Mean agent (both radiologist and model) confidence-accuracy correlation rose significantly, from a mean of 0.166 ± 0.112 without support, to 0.333 ± 0.116 with the model, affirming that both radiologists and models were significantly more likely to be confident in making a correct diagnosis, and vice versa, but that this was strengthened by support from the other agent ($p<0.05$). Mean agent confidence calibration (high confidence – low confidence accuracy rose significantly from 0.155 ± 0.138 without support to 0.333 ± 0.116 with support, indicating significantly more calibrated decision making with support from the other ($p<0.01$). Similarly, mean agent confidence bias (correct – incorrect confidence) rose significantly from 0.626 ± 0.437 without support to 1.420 ± 0.864 with support, indicating that, with support, radiologists and the model were significantly more confident when correct and less confident when incorrect ($p<0.01$).

In individual case-based assessment, radiologist confidence was significantly greater for correct predictions with model support (mean without model support 7.09 ± 1.89, mean with model support 7.61 ± 1.80, $p<0.0001$). Radiologist confidence was significantly lower in incorrect predictions with model support compared to without it (mean without model support 6.39 ± 1.92, mean with model support 6.12 ± 1.88, $p<0.05$). In short, when supported by a model, radiologists were more confident when making correct predictions, and less confident when they were incorrect.

Reviewing model confidence with and without radiologist support, the model was significantly less confident with radiologist support in both correct and incorrect predictions (mean model confidence in correct predictions without radiologist support 9.81 ± 1.02, mean with radiologist support 8.44 ± 2.12, $p<0.0001$; mean model confidence in incorrect predictions without radiologist support 10.00 ± 0.00, mean model confidence with radiologist support 6.67 ± 2.27, $p<0.0001$). Notably, for all incorrect model predictions, its confidence without support was at maximum (10/10), but was more calibrated with support from radiologists. Self-awareness and calibration quantisation showed that without agent support, only 2 of 12 agents (2/11 radiologists, and not the model) were within the optimal calibration quadrant. This rose significantly to 9 of 12 (8/11 radiologists and the model), with support from the other ($p<0.05$).

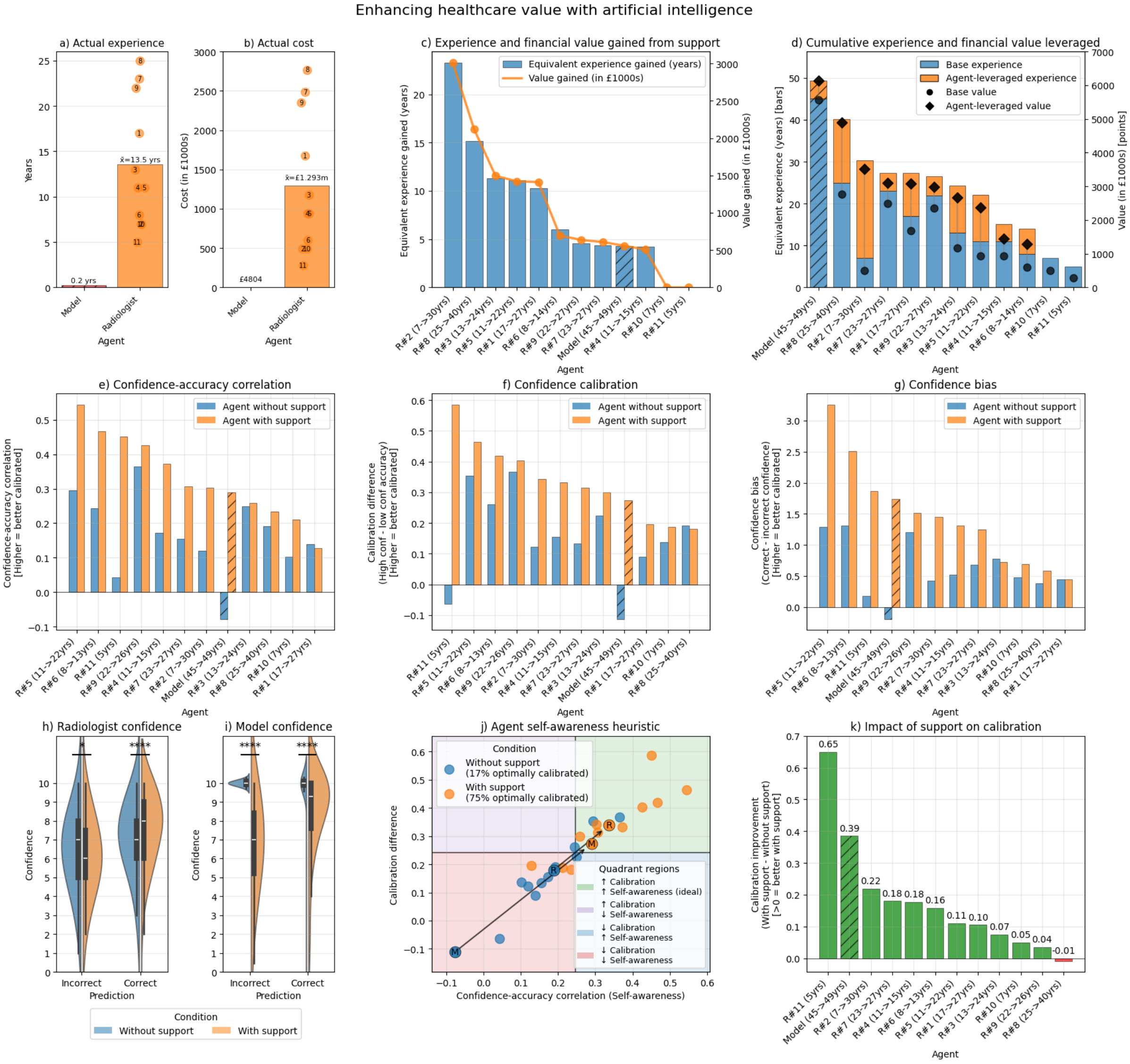

**Fig. 6. Enhancing healthcare value with artificial intelligence.** a) Model and radiologist experience. b) Model and radiologist training cost comparison. c) Additional experience (bars, in years) and financial value (line plot, in £1000s) gained in each agent leveraged with support. d) Stacked bar chart showing each agent's cumulative value, and additional value leveraged with support. e) Confidence-accuracy correlation for each agent with and without support. f) Confidence calibration showing the difference between high and low confidence accuracy with and without support; a higher value indicates better calibration. g) Confidence bias, showing the confidence bias for each agent with and without support, defined as the difference in confidence between correct and incorrect predictions; higher values mean agents are more confident when correct, and less confident when incorrect (better calibration). h) Violin plot showing radiologist confidence comparison between incorrect and correctly labelled images, with and without model support. i) Violin plot showing model confidence comparison between incorrect and correctly labelled images, with and without radiologist support. j) Agent self-awareness and calibration plot, showing confidence-accuracy correlation (self-awareness) on the x-axis, and calibration difference (accuracy with high confidence – accuracy with low confidence) on the y-axis. The higher calibration–higher self-awareness quadrant is ideal (the top-right, green-shaded area). Scatter points represent individual radiologists with and without model support, with 'R' representing the mean radiologist and calibration and 'M' representing the model. k) Model impact on agent confidence calibration, where more positive values denote an improvement in calibration with support.

## Discussion

Evaluating multiple radiologists, institutes, pathologies, and datasets, we demonstrate how clinical practice can be enhanced through partnerships between clinically trained human agents—in this case, board-certified radiologists—and trained AI agents. Whereas most contemporary clinical AI research evaluates performance gains from human agents supported by AI[17-20], here we expand upon that narrative to also assess the alternative paradigm: whether an AI supported by a human agent can leverage their clinical experience to achieve performance gains (Extended Data Fig. 1). From this, we reveal that not only are there circumstances in which an AI agent supported by a human can achieve performance gains, but rather that this paradigm may bring the greatest overall benefit. This applies not only to the individual patient but also across all stakeholders, including the patient, practitioner, and healthcare provider.

The study's niche is medical imaging, namely the domain of medicine in which the greatest strides in AI advancement have been made to date[48]. We evaluated performance on a task that is especially difficult: predicting whether a brain tumour would show contrast enhancement following intravenous contrast administration, but with human and AI agents permitted only to review the pre-contrast MRI sequences. From this, we identify that both a human agent (trained radiologist) supported by a trained AI agent and an AI agent supported by a radiologist improve their ability to identify lesional enhancement using the pre-contrast brain MRI alone. Not only does this translate to tangible improvements in classification accuracy, directly beneficial to the individual patient, but also to healthcare agents (both radiology staff members and the AI model), with increased reporting confidence and consistency. This, moreover, benefits healthcare systems, as we show that AI and human partnerships increase both the financial and experiential value of healthcare agents and can yield marked increases in human efficiency. Overall, our findings provide strong support for the applicability of AI to enhance existing diagnostic radiology services, including greater precision, faster turnaround times, reduced dependence on contrast use, and reduced healthcare disparities and reporting disagreement.

Pertinent to the specific task, these technologies (as investigated by many others elsewhere[24,42-47]) may also plausibly reduce dependence on intravenous contrast use, requisite for current post-contrast MRI acquisition. This could bring global environmental benefits by reducing the many thousands of litres of gadolinium administered, secreted and/or disposed of each year[49], a desirable outcome given that the long-term effects of gadolinium remain unknown, despite it now being detectable in sewage, surface, drinking water[50-52], water sources distant from MRI facilities[53], and even in fast-food soft drinks[54].

Our findings closely align with and extend the recent evidence on human-AI collaboration in medical imaging[17-20]. While many studies focus on AI's potential to improve diagnostic accuracy, most are more narrowly focused on single-beneficiary outcomes for either the patient or the provider, but not both[1-7]. Critically, our work seeks to address this tendency to optimise for single-stakeholder benefits, for which, across healthcare AI research, we strongly advocate assessing system-wide value creation. Notably, the Institute for Healthcare Improvement's quintuple framework puts forth the following aims for future healthcare improvement: 1) improving population health; 2) improving care; 3) advancing health equity; 4) improving workforce well-being, and 5) lowering cost[55]. But though widely adopted, this is rarely applied to AI evaluation. While AI tools can leverage large datasets to surpass human performance in specific tasks, their implementation often fails to account for the complex interplay among patient outcomes, clinician experience, and healthcare system sustainability. Our study uniquely demonstrates that when AI systems are designed with all three beneficiaries in mind, the cumulative value exceeds the sum of its parts.

Importantly, the results challenge prior research on chest x-ray reporting by Yu et al., who have recently suggested that human experience-based factors do not reliably predict the impact of AI assistance[19]. We would note that the task demanded of radiologists in our study, namely predicting whether a tumour will enhance in post-contrast imaging, but with only sight of pre-contrast sequences, is an especially challenging one that demands significant clinical training, often involving secondments to specialist centres both within the UK and even abroad to acquire the necessary experience (indeed, all of our study's participating radiologists had done this). This contrasts greatly with chest radiograph reporting for simple diagnostic assessments, such as pleural effusion, lung consolidation, and pneumothorax, which were employed in the comparator study, all of which are taught early in both medical school and non-medical allied health professional courses (e.g. nursing, radiography)[19]. Instead, we suggest that human experience-based factors may, in fact, reliably predict the impact of AI assistance when the task at hand is challenging and conventionally demands highly specialised levels of expertise.

The strengthening of the human agent triadic experience-confidence-accuracy relationship we illustrate with AI support suggests that AI does not merely add fixed performance boosts either. Rather, here we demonstrate that AI can help human agents better leverage their existing expertise to amplify accuracy, confidence, consistency and efficiency. Such models, therefore, have an instrumental role in healthcare delivery, and plausibly via metacognitive benefits.

The alternative approach is also revealing: an AI model supported by radiologists becomes more accurate and more calibrated in its predictive confidence. In short, regardless of whether an AI or a human agent is the primary 'driver' of a clinical task, we show that they improve their experiential and financial value by leveraging the experience of the alternate agent, including when a human supports an AI agent. Though this paradigm is more rarely studied clinically (with the focus instead on assessing the benefit of AI in supporting a human practitioner), these findings closely align with broader societal, industry, and academic advances in agentic AI, whether in autonomous driving[21], aviation autopilots[22], collaborative robots in manufacturing 'cobots'[56], or AI-assisted scientific discovery[57].

Our confidence calibration findings also provide critical insights into an underexplored dimension of the clinical human-AI partnership. The improvement in the confidence-accuracy correlation and the transition of both radiologists and the AI agent into the optimal calibration quadrant represent a fundamental improvement in safe and effective clinical decision-making. This addresses what Lee *et al*. previously identified as a major source of diagnostic errors: the misalignment between confidence and accuracy[41]. By improving calibration, we show that AI does not merely help radiologists make better decisions; it helps them know when they're making better decisions and, similarly, for the AI when supported by radiologists. This indicates metacognitive improvement in both man and machine, which holds profound implications for patient safety.

The healthcare system benefits we quantified—a median addition of six years of experience and £696,176 in additional value leveraged for each radiologist—provide empirical evidence of AI's potential to address the global radiologist shortage. With radiology facing a 30% workforce deficit in many countries[58,59], our findings suggest that an AI partnership could effectively expand capacity and, rather than compromise quality, enhance it. Moreover, our finding that, with the assistance of human agents, the AI model's estimated financial and experiential value rose by £559,528 and by 4 years also underscores an opportunity to evolve industry paradigms. Intuitively, it seems that to eke out the greatest benefit from an AI system, it should fully harness existing human domain expertise.

Our study has an array of limitations. First, regarding the patient data at our disposal, there was a sample imbalance, with fewer paediatric cases than adults included in the model-radiologist evaluation. This was beyond our control due to the limited sample sizes from the global datasets at our disposal. We opted for maximal patient inclusivity regardless of the sampling space to maximise equity and generalizability. Future studies should investigate the applicability of this approach to imbalanced samples more closely. Second, whilst the model we utilise is performant and, on many occasions, successfully completes what human radiologists may not, it is not perfect. AI model performances are ubiquitously asymptotic, and like all others, there is room for improvement, a task for future study. Third, it should be noted that the financial analysis presented is not exact but rather an inference drawn from contemporary clinical job contracts and salary scales. It could not reflect salaries from decades prior, nor could it anticipate salary costings decades in the future. Moreover, since the model value is derived from fitting a regressor to radiologists' experience and performance, we reiterate that these results are more exploratory and require dedicated health economics research to be fully evaluated. Finally, our expert radiologist sample is modest, but not large. Faced with the choice of recruiting a finite number of highly specialised individuals to evaluate a challenging clinical task versus a larger pool with more rudimentary experience undertaking a simpler one, we opted for the former, for we explicitly wished to investigate the benefit AI may bring to already highly specialised and trained individuals when faced with an especially difficult scenario. Given the study's demands, requiring each radiologist to review 100 unique cases twice, it was not feasible to recruit additional experts within the study window. Though we considered halving the caseload to recruit double the number of radiologists, we wanted to ensure a reasonable number of cases were drawn (by random selection) across all study sites and pathologies, hence our choice of evaluation. To our strength, our work includes 1100 radiologist-case evaluations analysed via a multi-case multi-reader randomised crossover trial (the contemporaneous gold standard[60,61]) and leans on the expertise of clinicians who provide frontline care across some of the UK's largest and most specialised hospitals.

In conclusion, we demonstrate that a human-AI partnership improves accuracy, confidence, and reporting speed for both human agents (trained radiologists) and AI agents when reviewing pre-contrast brain tumour MRIs to evaluate for possible post-contrast enhancing lesions. This applies to both partnership paradigms, whether AI supports a human or a human supports an AI. Though already shown in other areas of science and industry, but rarely studied clinically, the greatest overall benefit to patients, practitioners, and healthcare providers was, in fact, evident with an AI agent supported by a human expert. The study provides compelling evidence that such partnerships simultaneously benefit patients, practitioners, and healthcare systems. Synergistic improvements in agent accuracy, confidence calibration, and inter-rater agreement suggest that AI's greatest contribution may be to create more capable, confident, and consistent clinical agents, whether human or model-based. Our work highlights that the maximal AI value in healthcare could emerge not from replacing human intelligence, but from AI agents leveraging and amplifying it. Future AI development in healthcare should consider both paradigms of the human-AI partnership, including across all plausible benefits to the patient, practitioner, and healthcare provider.

## Methods

### Study design

We conducted a multi-case, multi-reader, randomised crossover expert agent performance study, following established guidelines for diagnostic accuracy studies (STARD). Here, the expert agent under evaluation could be either a human radiologist or an AI agent trained to detect and segment post-contrast-enhancing brain tumours from only non-contrast imaging sequences. The study employed a within-reader design, in which each radiologist and the model served as their own control by evaluating cases both with and without the assistance of the other corresponding agent (Extended Data Fig. 1, Extended Data Fig. 2, Extended Data Fig. 3).

### Neuro-oncology patient cohort

We utilised MRI brain imaging data from a diverse cohort of 11,089 unique adult and paediatric patients with brain tumours. These patients had a wide variety of histopathologically-confirmed diagnoses, including subtypes of glioma, meningioma, metastases (from various primary malignancies), paediatric lesions, and post-surgical resection cases. These data were drawn from multiple international sites, including the UK[42,62], the USA (US-based BraTS challenges[63-67], UCSF-PDGM[68], UPenn-GBM[69]), the Netherlands[70], and Sub-Saharan Africa[71]. From this pool, we extracted the 1109 held-out test-set cases from the deep learning model's prior development[42], ensuring no possibility of information leak to its training set. This cohort was as maximally heterogeneous and inclusive as could be achieved, spanning both adult and paediatric cases, pre- and post-operative imaging, multiple neuro-oncological pathologies, and numerous global study sites.

Images were pre-processed using well-established, validated techniques published by others elsewhere. In brief, for internal data, this included automated sequence labelling[72], intensity clamping (to attenuate signal artefacts), super-resolution[73], and brain extraction[74]. For all external cohorts, sequences were pre-labelled, pre-processed, and many were already brain-extracted. Ground truth-enhancing tumour masks were always generated from T1-weighted, T2-weighted, FLAIR-weighted, and post-contrast T1-weighted sequences (i.e., complete datasets), which were available for all cases under study. For external datasets, this was already available in the source data, as described above. For local data, we used an open-source tumour segmentation model, applied to complete datasets in which post-contrast imaging was always available, to derive labels for enhancing tumour, non-enhancing tumour, and perilesional signal change[24]. Inferred labels were reviewed before their downstream application by a neuroradiologist with seven years of neuro-oncology experience. Having segmented lesions in the patient's native space, structural and lesion segmentation masks were nonlinearly registered to 1x1x1mm isotropic MNI space with Statistical Parametric Mapping (SPM, v12) and enantiomorphic correction[75,76]. The advantage of enantiomorphic correction is that the risks of registration errors secondary to a lesion are minimised by leveraging a given patient's normal structural neuroanatomy on the unaffected contralesional hemisphere[75].

## An AI model to segment contrast-enhancing brain tumours using only non-contrast MRI sequences

We applied a previously trained and validated deep learning model for the cardinal task of identifying and segmenting enhancing tumour using only pre-contrast MRI sequences (namely T1-weighted, T2-weighted, and FLAIR-weighted). The development of this model is described in significant detail elsewhere[42]. In brief, the model was trained on a cohort of 9980 patient samples across the pathologies, cohorts, and countries as described above, where ground-truth enhancing tumours were pre-determined using harmonised segmentations from expert radiologists and segmentation models trained on complete datasets (namely, including a post-contrast T1-weighted sequence). The model architecture was nnU-Net, trained with 4000 epochs, with the large residual encoder adaptation[77,78], where its task was to segment normal brain, abnormal tissue, and enhancing tumour (the latter being the primary interest) with access only to the pre-contrast T1-weighted, T2-weighted, and FLAIR-weighted sequences. This was the primary aim of assessing whether such models could reduce contrast dependence in patients in whom repeated gadolinium administration may be undesirable, including but not limited to those allergic, in renal failure, undergoing extensive imaging follow-up, or paediatric populations. None of the cases used in this article's analysis were used in the model's training set.

## Assessing model-assisted human agents

We invited UK board-certified (post-Fellowship of the Royal College of Radiologists [FRCR], post-Certificate of Completion of Training [CCT]) radiologists to participate in a case-review study. Each radiologist was assigned 100 cases from the 1109 neuro-oncology patient cohort. Within this 100-patient reporting allocation, 50 were randomly drawn from samples that did contain contrast-enhancing tumour, with the remaining 50 randomly drawn from the samples that did not (i.e. showed only nonenhancing tumour, or a total resection with no contrast-enhancing residuum).

We subsequently undertook a crossover trial of radiologists reviewing the pre-contrast imaging, evaluating their decision-making as to whether gadolinium should be given for post-contrast sequences (Extended Data Fig. 3). Each radiologist sequentially reviewed each case and was first shown the non-contrast MRI sequences (T1, T2, and FLAIR) alone, and, on a second occasion (within the same week), was shown both the non-contrast sequences and the model's prediction of whether there would be post-contrast enhancing disease. In short, each radiologist would review a given case twice, enabling parallel comparison, but would never be shown the post-contrast images. Radiologists reviewed these cases on a Linux workstation with a 4K monitor using ITKSnap, all of whom were trained to use the software. For each case, the radiologist was asked the following questions: 1) *Do you think there will be an enhancing abnormality in this case?* (answered as yes or no); 2) *How confident are you*? (rated via Likert scale between 1 and 10); and 3) *What is your assessment of image quality? (*rated via Likert scale between 1 and 10*).* All received standardised training on the review platform. We also recorded each radiologist's prior years of experience and the time they required to review each case. Cases were presented in a randomised order to minimise recall bias. Radiologists were always shown the 100-case tranche without the model segmentation first, since it was deemed plausible that they might otherwise recall (and therefore be biased by) the model's prediction in future assessments.

## Assessing human-assisted AI agents

We also considered how incorporating human expert priors and assessments may augment AI agent decision-making. To evaluate whether incorporating radiologist assessments could improve AI performance, we developed and optimised a calibrated integration strategy. We extracted the radiologist-reviewed subset of cases (n=1100) and performed hyperparameter optimisation using nested cross-validation to prevent optimistic bias from data leakage. The integration strategy combined model predictions with radiologist assessments using three key parameters: (i) selective enhancement mode (selective application based on model uncertainty), (ii) human weighting factor (0.1-0.8, controlling the relative contribution of radiologist input), and (iii) model uncertainty range (threshold for triggering human input integration, 0.4-0.6). Model confidence was quantified from segmentation probability maps, with binary detection defined as a Dice coefficient >0.3. Radiologist assessments included confidence scores (1-10 Likert scale) and binary enhancement predictions. We employed a rigorous nested cross-validation framework with five outer folds for unbiased performance estimation and three inner folds for hyperparameter selection. This double cross-validation approach ensures that (1) hyperparameters are selected independently of test data in each outer fold, and (2) performance metrics reflect true generalisation rather than overfitting to the validation set. To ensure robustness and stability of both performance estimates and optimal hyperparameters, we repeated the entire nested cross-validation procedure with five different random seeds.

## Statistical analysis

Our primary outcomes were to determine whether there was any change in a radiologist's reporting accuracy with model assistance and, conversely, whether there was any change in a model's reporting accuracy with radiologist assistance. Our secondary outcomes were to assess changes in confidence, response time, and inter-rater agreement, and the relationship between these features with and without support from the other agent.

All analyses were performed and reported in accordance with the TRIPOD and PROBAST-AI guidelines[79]. For the overall radiologist AUC, sensitivity, and specificity, both with and without model support, we undertook a multi-case, multi-reader analysis using MRMCaov[80], the established gold-standard approach[60,61], which allows control of both case difficulty and within-reader performance differences. The analysis of additional metric differences was primarily conducted at the agent level, reporting reader-averaged accuracy, precision, recall, and F1 scores, also in accordance with common standard practice[81].

Statistical tests included the Chi-squared test, McNemar's test for paired binary comparisons, paired and independent samples t-tests, Mann-Whitney U test, Fisher's exact test, and one-way ANOVA with Tukey HSD post-hoc tests. Inter-rater agreements were assessed by Cohen's kappa. ROC and PR curves were derived from radiologist performance and confidence, as well as model performance and probability map scores (the best available proxy for confidence). Linear and logistic mixed-effects regression models were used to fit models comparing radiologist accuracy, reporting speed, confidence, and years of experience while accounting for the reporting radiologist. Model comparisons were performed using likelihood ratio tests, with AIC and BIC metrics used for model selection. Pearson and Spearman correlation coefficients were used to assess relationships between continuous variables as appropriate. Variance homogeneity was assessed using Levene's test, and variance ratios were compared using F-tests with bootstrapped confidence intervals. UMAP was used to embed the three-dimensional ground truth masks into a compact two-dimensional space, from which the relationship between lesion size, shape, and morphology could be compared to these outcomes. Within domain statistical tests were corrected for multiple comparisons using Bonferroni correction within hypothesis families[82].

Combining model fits with both a radiologist's years of experience and the UK doctor pay scale[83], we extracted the financial and experiential value added (if any). To do this, we extracted the slopes from the two model fits: 1) $Accuracy \sim \beta_0 + \beta_1(Years\ of\ experience)$ and 2) $Confidence \sim \beta_0 + \beta_1(Years\ of\ experience)$. We also calculated the accuracy and the gain/loss in confidence within each human agent with model assistance. Using the fitted slopes, we then derived the equivalent accuracy and confidence when supported by the model and averaged the two. Similarly, we fitted the model to this function, drawing on its accuracy and confidence scores, to infer a base experiential and financial value, fitted to these same human radiologist workforce and pay schedules[83]. This enabled us to infer the model's experience and value in isolation, and how that changed when supported by radiologists in its decision-making.

We also quantified the confidence calibration and confidence bias in both human agents and the AI model. Firstly, confidence calibration was assessed by identifying the 25$^{th}$ and 75$^{th}$ percentiles of confidence scores with and without agent (model or human) support. We then derived the corresponding high and low confidence accuracies (i.e., the percentage of correct reports when confidence was either ≥ 75th percentile or ≤ 25th percentile). We quantified the calibration difference as the difference between high-confidence accuracy and low-confidence accuracy, where positive values would indicate good calibration (i.e. high confidence relating to high accuracy) and near-zero or negative values would indicate either poor calibration (i.e. no relationship between reporting confidence and accuracy) or inverse calibration (overconfidence with low accuracy). Similarly, we also assessed the confidence bias across all agents, extracting the mean confidence on correctly and incorrectly labelled cases. Confidence bias was then defined as the difference between the two, with higher values indicating better calibration (more confidence when correct), values tending to zero indicating poor calibration (equal confidence regardless of reporting accuracy), and negative values indicating inverse calibration (more confidence when incorrect). By extension, we inspected the relationship between radiologist and model self-awareness and confidence calibration, reviewing the relationship between calibration difference and self-awareness, quantified as the Pearson correlation coefficient between confidence score and accuracy within individual scan reviews, across all radiologists and the model, across both conditions. This yields a calibration plot of four quadrants (split by median values) of 1) high self-awareness and good calibration (ideal), 2) poor self-awareness but good calibration, 3) good self-awareness but poor calibration, and 4) poor self-awareness and poor calibration. We used Fisher's exact test to statistically test whether quadrant allocation differed for radiologists with and without model support, and similarly, whether quadrant allocation differed for the model with and without radiologist support.

## Data availability

Most of the patient data utilised in this article is freely and publicly available[63-71]. Our study ethics preclude the release of internal data.

## Code availability

Openly available Python software and models were used for all model development and downstream analyses[24,80,84-91].

## Ethical approval

All external datasets were granted ethical approval as described within the source datasets[63,65-71]. Descriptions of some of the imaging datasets used have been previously reported[24,39,63,65-71]. The local ethics committee approved consentless use of irrevocably anonymised data under standard opt-out frameworks.

# Bidirectional human-AI collaboration in brain tumour assessments improves both expert human and AI agent performance

Supplementary Material

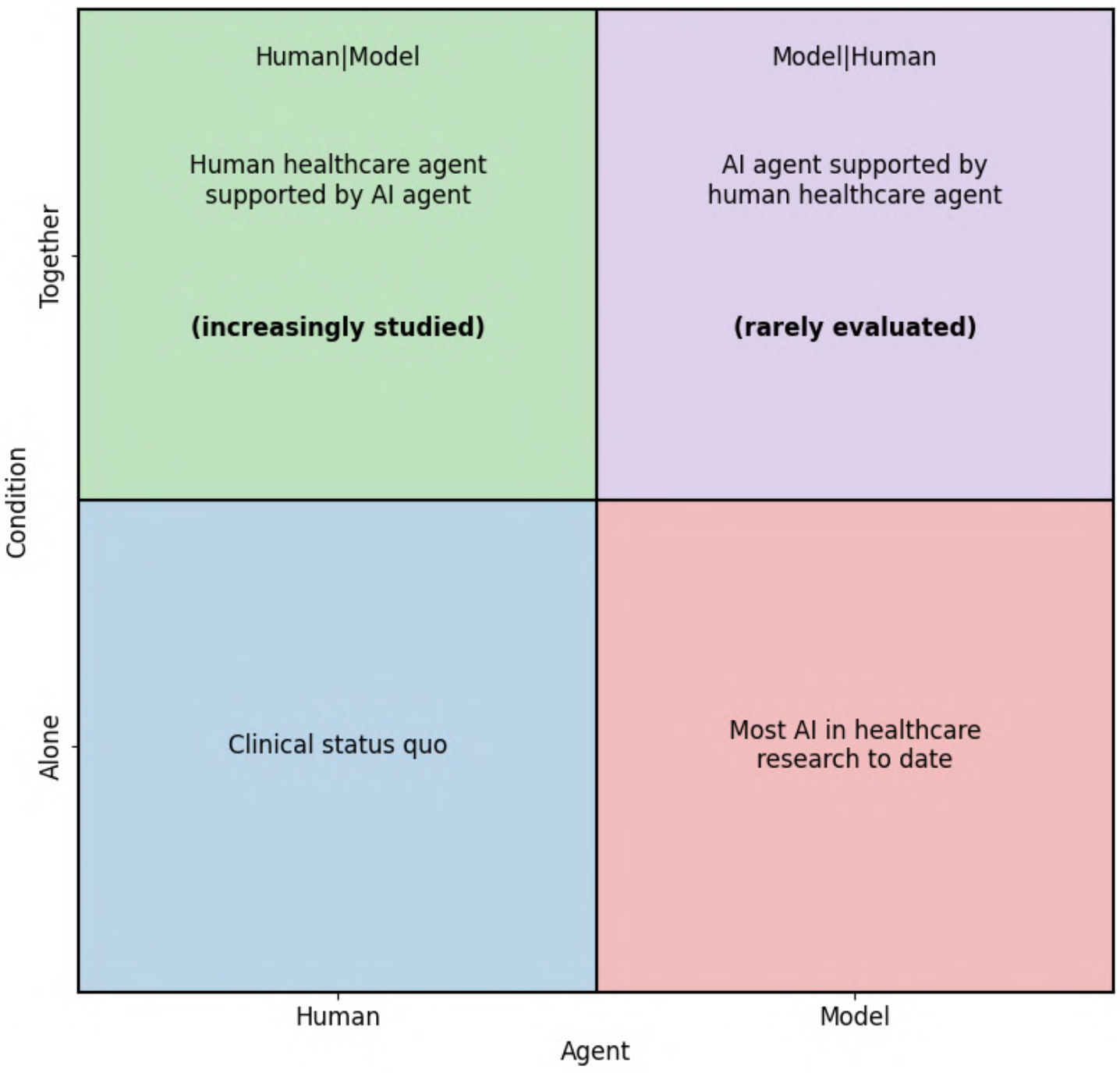

**Extended Data Fig. 1. Paradigms of evaluating AI value in healthcare.** Most previous AI in healthcare research has sought to compare performance gains from AI automation (red box) with respect to the clinical status quo, where it was ordinarily delivered by a human agent alone (blue). Increasingly studied approaches focus on unidirectional AI-human partnerships, evaluating how AI agents enhance human performance (green box). The alternate paradigm, how a human agent can assist AI agents (purple box), is, however, rarely studied.

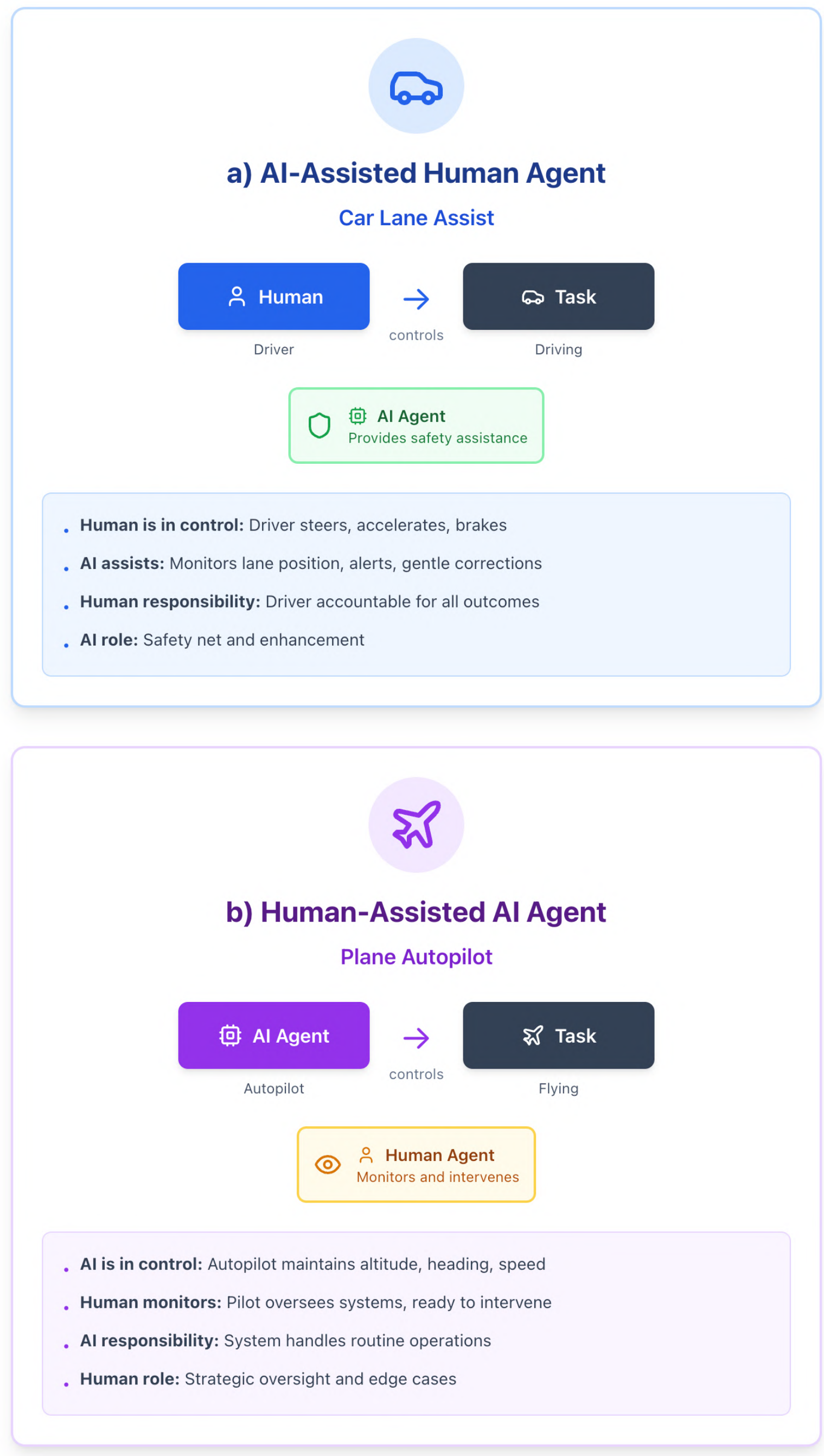

**Extended Data Fig. 2. Human-AI collaboration paradigms.** a) Most contemporary approaches to AI algorithm evaluation are those in which the model's utility is related to assistance with human performance: given the supplementation of an AI agent, how does the performance of human agents change? For example, are cars driven more safely by humans with AI-assistance tools, such as automatic lane-keeping? b) Though rarely assessed in medical research, the alternative paradigm is argued to be equally important: given the supplementation of a trained human agent, how does the performance of an AI agent change? For example, autopilot systems that undertake routine piloting operations, with a human pilot overseeing and intervening if or when required. This particular figure was generated via paradigm b.

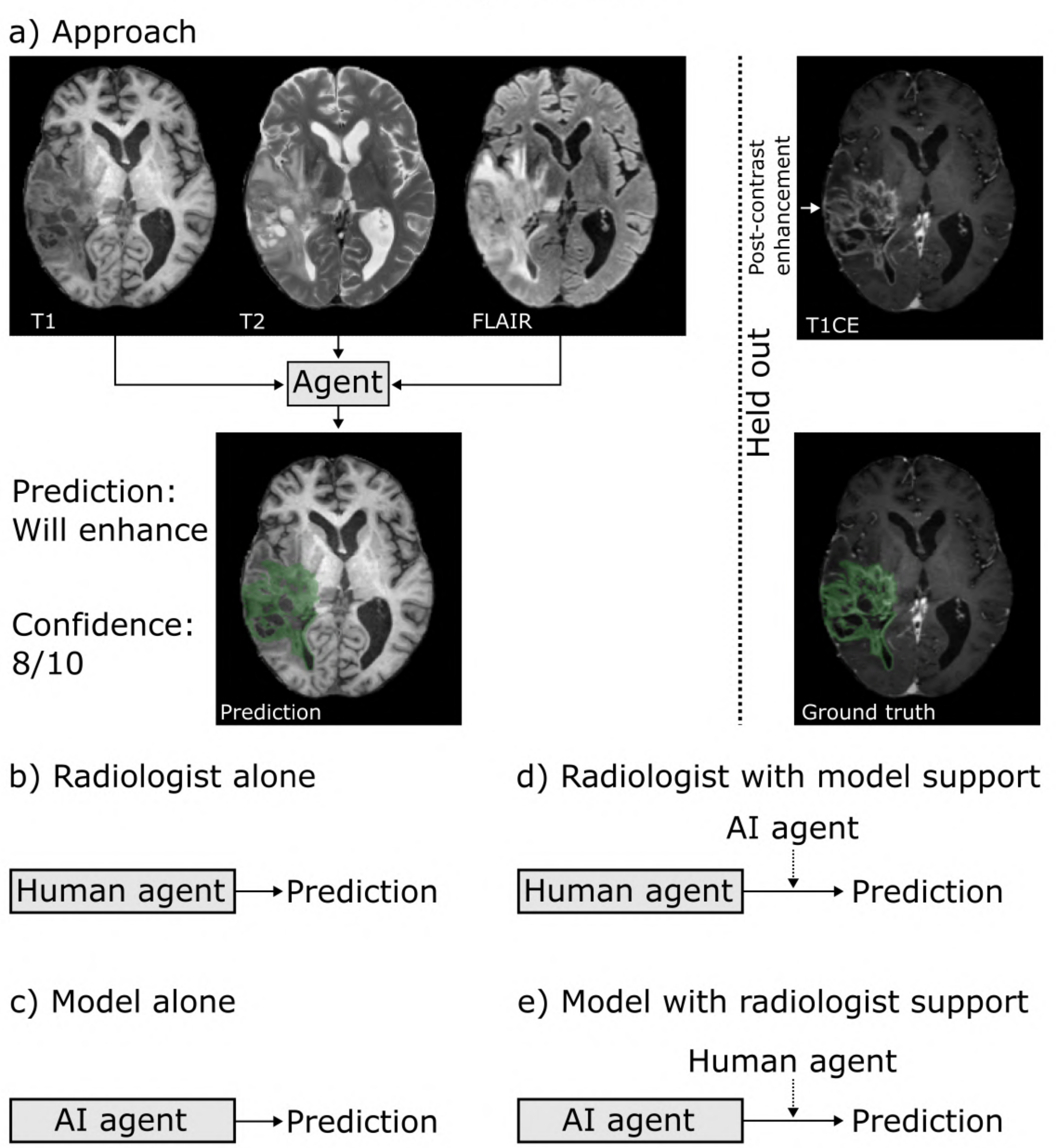

**Extended Data Fig. 3. Study schematic.** a) Experimental approach. Firstly, a previously trained human or AI agent was tasked with classifying whether an MRI scan of a patient with a known history of a brain tumour would or would not contain post-contrast-enhancing tumour. Post-contrast-enhancing tumour is ordinarily only detectable with the supplementation of the intravenously injected MRI contrast agent, gadolinium (post-contrast T1-weighted sequence [T1CE], as shown above). However, in this experimental design, the post-contrast T1-weighted image was withheld from both human and AI agent view as a means to yield a substantially harder task for both human and AI agents. b-c) We next evaluated the performance of b) human agents alone [trained radiologists], and an AI agent alone [trained model] in predicting the presence of absence of contrast-enhancing tumour from the non-contrast T1, T2, and FLAIR-weighted MRI sequences. d-e) Subsequently, we evaluated the performance of d) human agents supported by the AI agent, and e) the AI agent supported by human agents, in the same task.

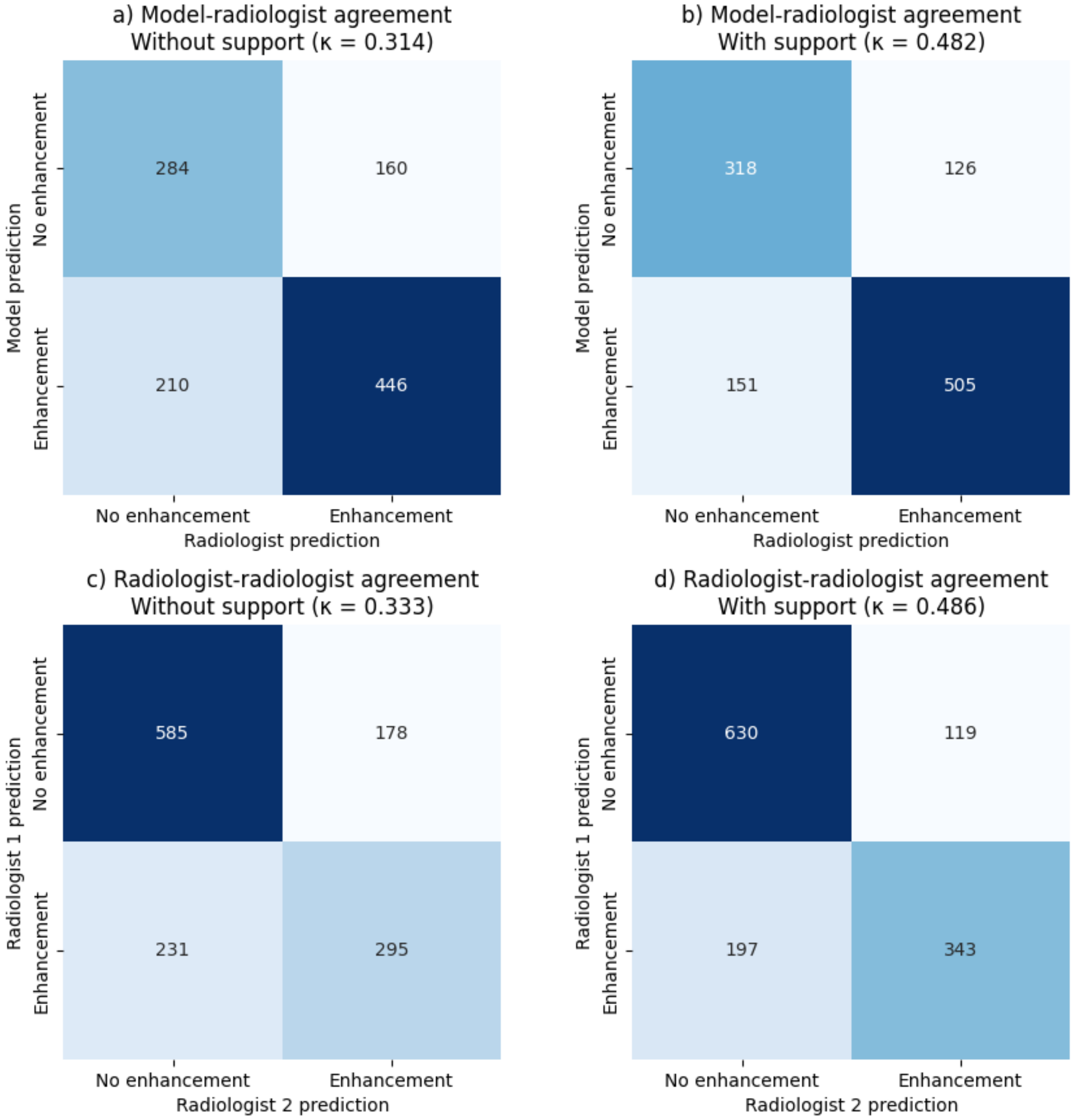

**Extended Data Fig. 4. Agreement comparisons.** a-b) Agreement between human agents (trained radiologists) and AI agents (trained model) with and without the support of the other. c-d) Agreement between the radiologists with and without the support of the model.

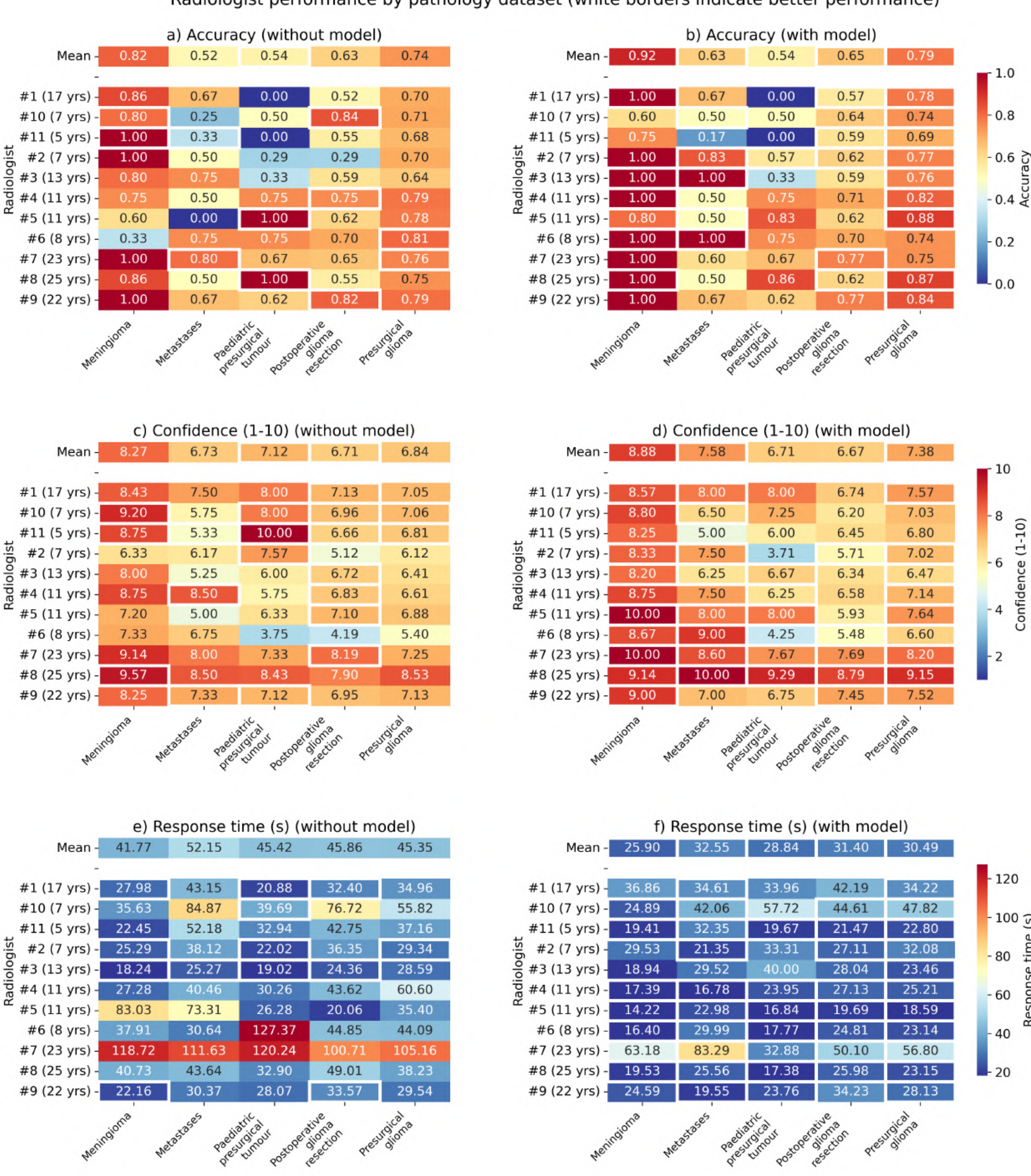

**Extended Data Fig. 5. Human agent performance by pathology dataset**. a-b) Individual radiologist accuracy with and without the model, c-d) confidence, and e-f) response times. For each heatmap, the x-axis denotes the different pathology classes, and the y-axis denotes each radiologist. White-box annotations around each cell indicate whether a given performance metric was better with or without model support.

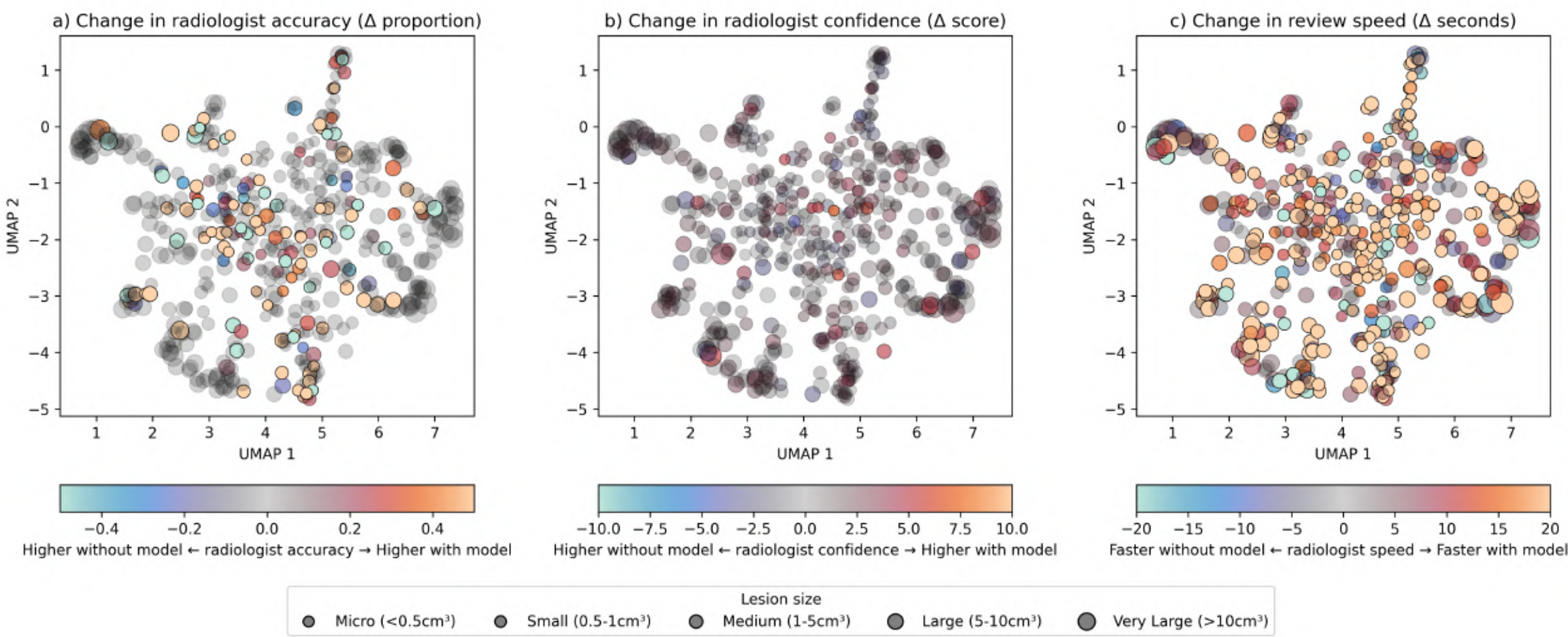

**Extended Data Fig. 6. Effect of lesion morphology, pathology, size, and distribution on human agent performance.** Panels show the UMAP representations of the lesion ground truth masks compacted into a two-dimensional space, where point size reflects lesion size and colour refers to the change in radiologist accuracy (a), confidence (b), and review speed (c).

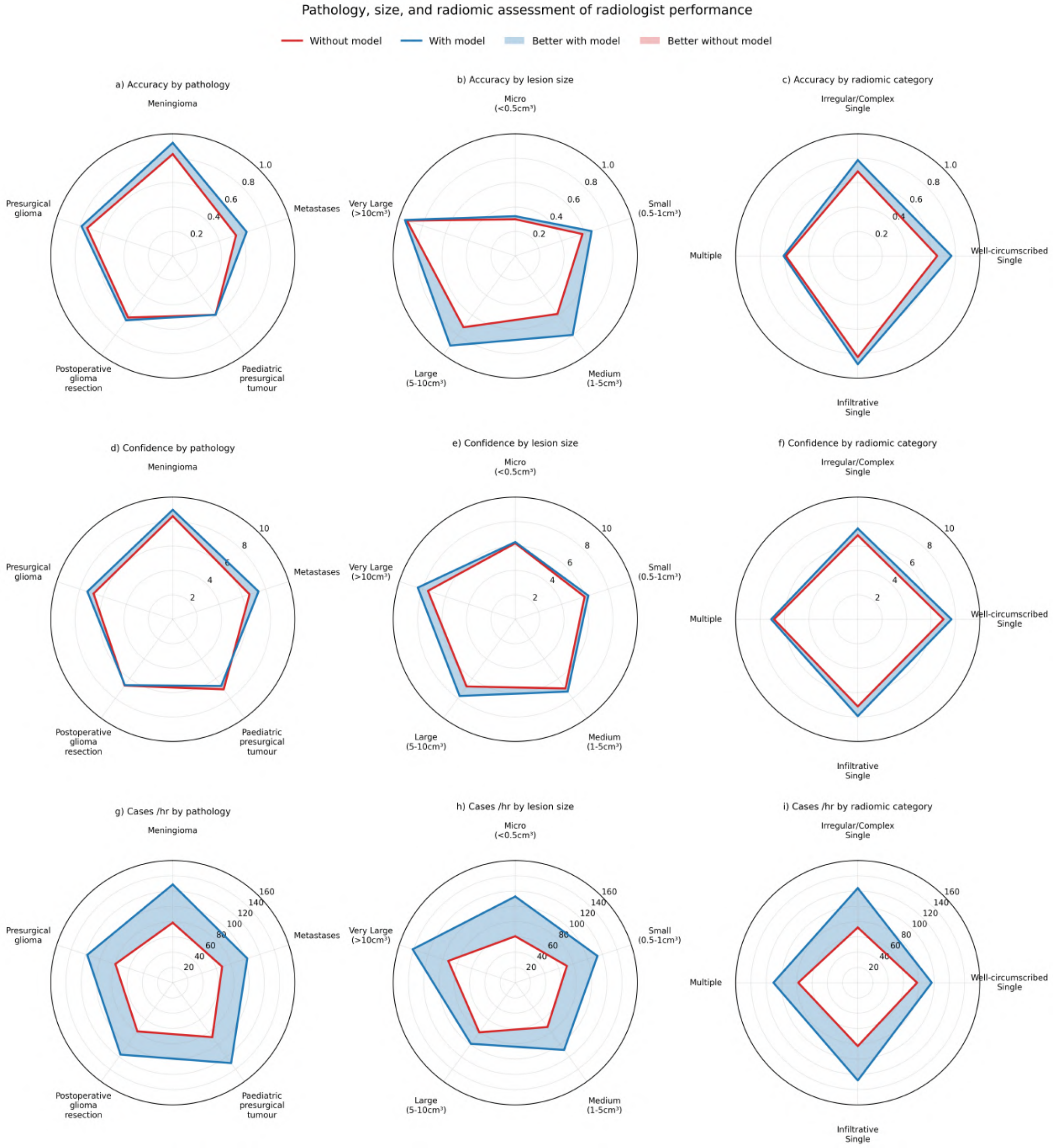

**Extended Data Fig. 7. Pathology, size, and radiomic assessment of radiologist performance**. a-c) Radiologist accuracy by pathology (a), lesion size (b), and radiomic category (c). d-f) Radiologist confidence by pathology (d), lesion size (e), and radiomic category (f). g-i) Radiologist throughput (in cases per hour), by pathology (g), lesion size (h), and radiomic category (i). Red lines indicate performance without the model, and blue lines indicate performance with it. Blue shading indicates performance gains with the model, and red shading indicates performance gains without it.